# On the helical behavior of turbulence in the ship wake

E. Golbraikh[1], A. Eidelman[2], A. Soloviev[3]

[1]*Physics Department, Ben-Gurion University of the Negev, Israel*

[2]*Mechanical Engineering Department, Ben-Gurion University of the Negev, Israel*

[3]*Oceanographic Center, NOVA SouthEastern University, USA*

## Abstract

Turbulent ship wake conservation at a long distance is one of unsolved problems at present. It is well known that wakes have a rotational structure and slowly expand with distance. Nevertheless, experimental data on their structure and properties are not sufficient. On the other hand, these experimental data show that the divergence of wakes does not change according to the law 1/5, as predicted by the theory.

In our work we study the effect of helicity on the parameters of a turbulent ship wake. Taking into account the helical nature of the wake, we can clarify the difference between turbulence inside and outside of the wake on the one hand, and slow its expansion with time.

## Introduction

A wake observed behind a ship on the ocean surface under relatively quiet wind conditions possesses some universal properties. In fact, the main wake structure includes, probably, two interacting zones, namely, Kelvin's wake and a turbulent wake retained over large distances with a weak angular divergence. Experimentally observed universality of the wake structure indicates to the universality and maybe self-similarity of the processes in it. A sufficient amount of numerical models have been suggested to describe the flow around a moving ship and in its vicinity, which satisfactorily reflect, to the first approximation, the processes in this zone. Regarding the wake, some of its features (e.g., Kelvin waves) have been studied extensively enough (see, for example, Reed and Milgram, 1993a, b; 2002; Zilman et al., 2004 and references therein). On the other hand, the central turbulent zone of the wake, well-observed (see, for example, Reed et al., 2002; Munk et al., 1987 and references therein) within sufficiently large distances in some field experiments, remains poorly studied as yet, and the description of its properties is far from being complete.

Dark-streak images, which can be several kilometers long under low wind conditions (2.5 to 7.5 m/s), are shown in Fig. 1. Measurements of short waves suppression in ship wakes were conducted and compared with SAR images of similar wakes in sevaral studies (Munk et al., 1987; Milgram et al., 1993a, b; Reed et al., 2002; Shen et al., 2002 ). They have shown that the phenomenon can be explained by the suppression of short sea waves in the wake, said waves being responsible for radar backscattering associated with a brighter background.

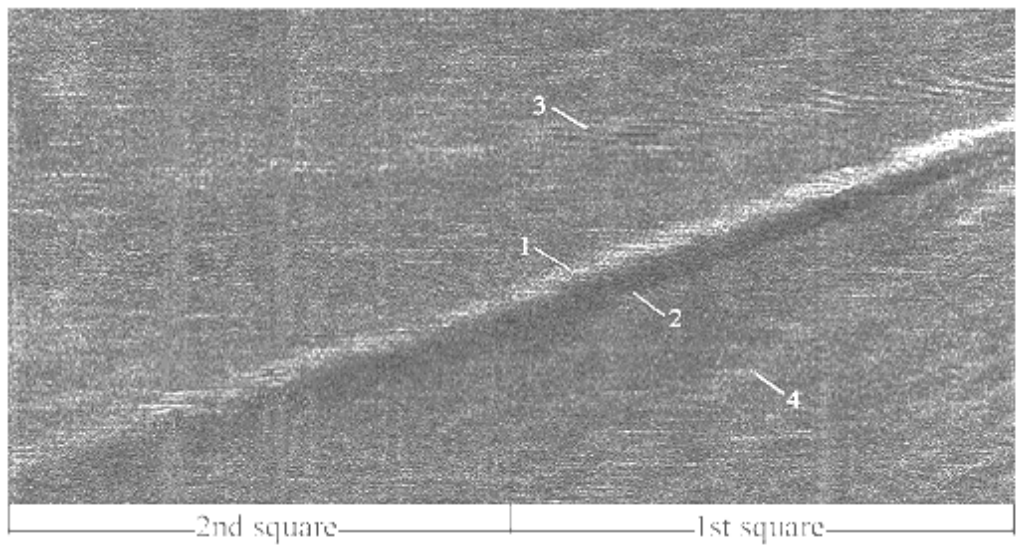

Fig.1 Ship wake

1. One of the arms of a narrow V-wake; 2 – turbulent wake;
3, 4 – boundaries of the Kelvin wake (after Zilman et al., 2004).

It remains unclear until now what physical processes occurring in the wake affect its properties, how initial (originated in the near zone) vortical disturbances are developed, how a system of large-scale vortices arises, how energy is supplied therein, etc. One of the main physical problems actively discussed at present is that of motion sources in far wakes leading to their long-living character. In the present paper, we do not discuss energy sources maintaining turbulent wake far away from a ship. Note only that they can comprise different waves generated on the water surface (for example, Kelvin's or wind waves). We dwell on turbulent processes leading to wake conservation or, in other words, on the processes decreasing its turbulent dissipation.

## Spectral characteristics of turbulent wakes

Experimental data on turbulent characteristics of surface ship wakes are limited. Only in few laboratory experiments (Patel, Sarda, 1990, Hoekstra, Ligtelijn, 1991, Benilov et al., 2000, Shen, 2002) certain turbulent parameters of the wake, including spectral characteristics, were measured. The ship-wake turbulence is still well detectable even in the far wake, where the Kolmogorov inertial range can be identified (Benilov et al., 2000). A sufficiently intense turbulence with Reynolds numbers corresponding to a developed turbulence is observed in the far wake. Benilov et al. (2000) explained the long-living wake structure by the presence of intense turbulence against the background of a shear flow supported by the energy of dissipated wind waves. From the standpoint of turbulent approach, it is generally assumed that turbulence is an energy sink for large-scale motions.

Wake velocity spectra obtained in the experiments of Benilov et al. (2000) at a distance x/L = 8.125 aft the model at the velocity Vs=126 cm/s are shown in Fig. 2, where our analyses of the spectra revealed a spectral slope close to –7/3. The spectral slope –7/3 is a characteristic slope of turbulence with a nonzero mean helicity (Brissaud et al., 1973). Mean helicity generation occurs in different turbulent flows with violated mirror symmetry, although turbulence is uniform and isotropic (e.g. Branover et al., 1999).

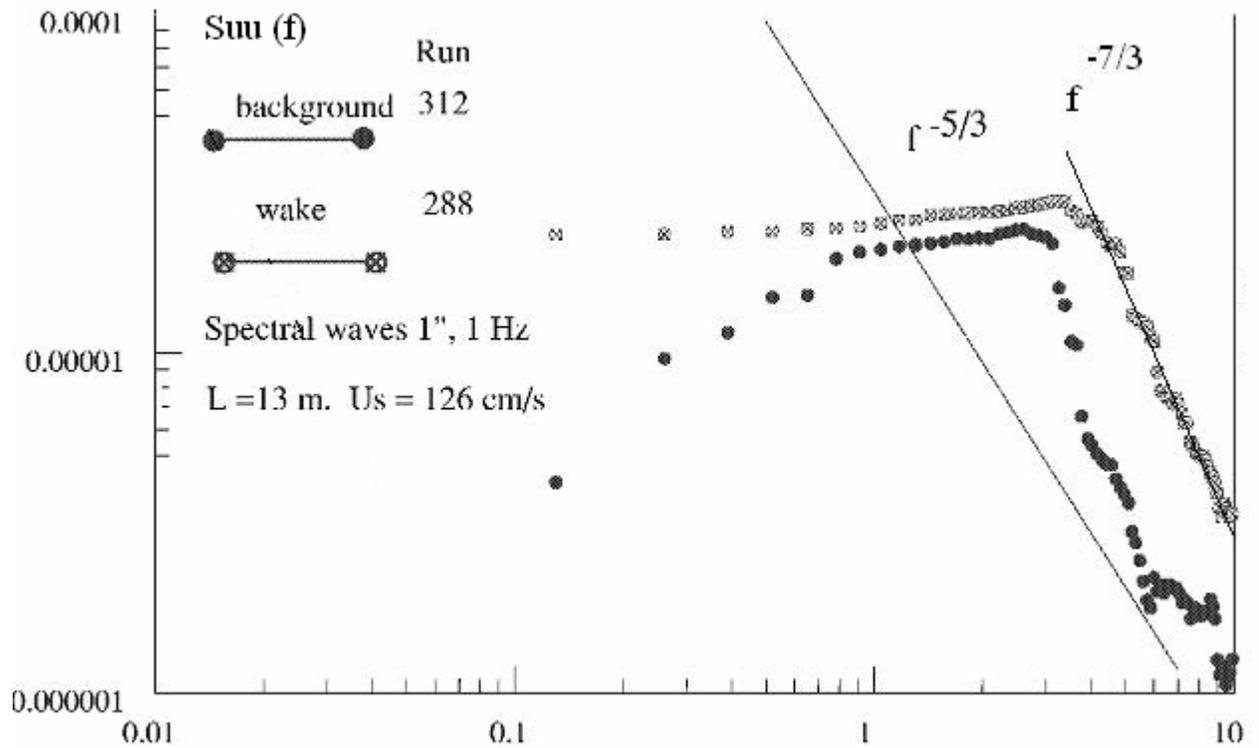

Fig.2 Wake turbulence under spectral waves conditions at a distance L/D=8.125 from the model, with the speed 126 cm/s, wave elevation variance 2.54 cm, spectral peak frequency 1 Hz (after Benilov et al., 2000).

Flow rotation observed in a wake leads, among others factors, to mirror symmetry violation of turbulence. In this case, even a uniform and isotropic turbulence acquires additional properties, so that a nonzero Eulerian integral of motion – mean helicity $He = \langle \vec{u}'\cdot curl\vec{u}' \rangle \neq 0$ appears in it. Here $\vec{u}'$ is a turbulent component of the velocity field, and <…> denotes averaging over the ensemble. Mean helicity, along with energy, is an essential invariant characteristic of a turbulent flow. However, a mean helicity value is often too small to affect considerably the flow behavior. One can hardly expect that a motion supported by pressure gradient only, in the absence of boundaries, would possess helical properties.

Mean helicity generation in the presence of rotation is a highly probable and important natural reason for the formation and conservation of the large-scale longitudinal vortical structures of ship wakes. The possibility of mean helicity origination in turbulent rotating flows was discussed in detail by Krause and Rädler (1980). As shown by the authors of this work, in the presence of differential rotation and/or external forces, helicity reaches appreciable values and can affect flow characteristics. As noted by Benilov et al. (2000), the flow structure in the turbulent wake can be rated as an ordinary shear, and therefore, we can speak with certainty about a relatively high helicity in a turbulent wake (Chkhetiani et al, 1994).

We examine some peculiarities of spectra in the presence of helicity. Spectral characteristics of helical turbulence were first studied by Brissaud et al. (1973). Here the helical spectrum of the spectral density of turbulent energy $E(k) \propto \eta^{2/3} k^{-7/3}$ (where $\eta$ is the helicity flux and $k$ is the wave vector) appears along with the Kolmogorov's spectrum $E(k) \propto \varepsilon^{2/3} k^{-5/3}$ (where $\varepsilon$ is the energy flux). According to the approach of Brissaud et al. (1973), helical spectrum should exist only for $\varepsilon = 0$. However, as follows from experimental data (Manson and King, 1985; Nastrom et al., 1987; Lilly and Petersen, 1983; see also Branover et al., 1999 and references therein), the helical spectrum is not at all exotic, being observed in various flows, and in these flows the energy flux is also nonzero.

Both slopes are also observed in other spectra presented by Benilov et al. (2000), e.g. shown in Fig. 3 for the wake cross-section at the same distance and at a velocity Vs = 202 cm/s. A large-scale range of spectra is close to –7/3 slope and an adjacent small-scale range – to –5/3 slope.

Generalizing previously obtained results, turbulent energy density spectrum has been analyzed using Kolmogorov's approach for the inertial interval within the framework of the asymptotic model in the case where ε and η are governing parameters not only in the inertial interval (Golbraikh and Eidelman, 2007). They have shown that the properties of the structure function in the inertial interval are affected by the behavior of the determining parameters ε and η in adjacent regions, both large-scale and dissipative ones. In such a case, the spectra become complicated, and their slopes defined by mutual influence of the regions can arise (see Fig. 3). Therefore, –7/3 slope of spectra points to a noticeable nonzero mean helicity.

Basic properties of helical turbulence having a nonzero mean helicity are an enhancement of large-scale helical vortical structures and a decrease in turbulent diffusivity (see Moiseev et al., 1983a, b; Belyan et al., 1993; Belyan et al., 1998; Branover et al., 1999 and references therein). Note that helical turbulence generates and/or enhances coherent structures in turbulent flows. When speaking about generational properties of turbulence, we imply that a turbulent cascade differs from Kolmogorov's one, where the energy supplied from the outside into large-scale fluctuations is directly transferred into small scales and dissipates therein. It can lead to a change in the spectral behavior of turbulent fluctuations connected with the mean helicity level of the turbulent field.

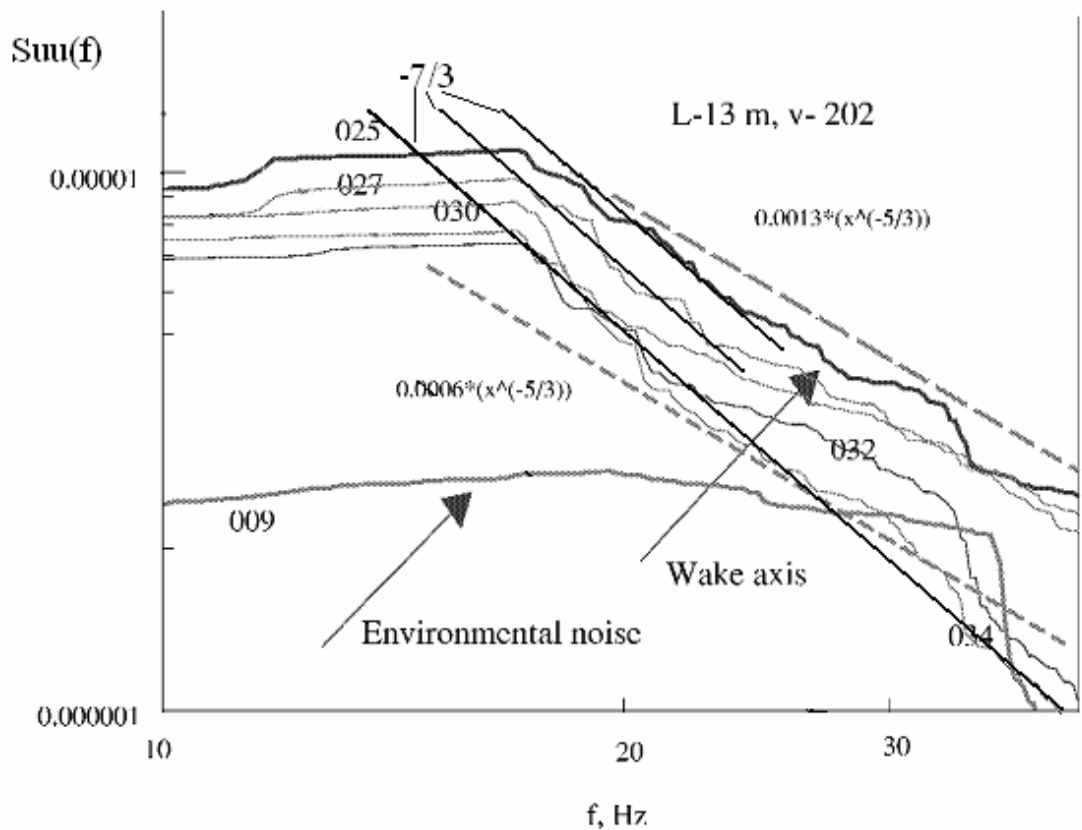

Fig. 3 Turbulent spectra in the wake cross-section at the distance L = 13 m from the model. Model speed Vs=202 cm/s (after Benilov, 2000)

Helicity level in a far wake (~ 21km aft a ship) can be estimated using field experimental data of Peltzer et al. (1992), where the characteristic vorticity scale $L \sim 20m$. It is assumed that two vortices coexist in the far wake and that turbulence is mainly concentrated within 0.5-0.6 of the vortex radius (Patel and Sarda, 1990). Main vorticity fluctuations produced with fluctuations of lateral velocity are located at a certain distance from the surface (Shen et al., 2002) and amount to

$\Omega \sim 0.05 - 0.1 / L, s^{-1}$. The estimated helicity flux reads $\eta \sim 3 \cdot 10^{-8} \div 1.3 \cdot 10^{-7} m / s^3$, and the estimated turbulent energy flux amounts to $\varepsilon \sim 6 \cdot 10^{-6} - 5 \cdot 10^{-5} m^2 / s^3$. On the other hand, under experimental conditions (Peltzer et al., 1992) with the wind speed not exceeding 5 m/s, the value of $\varepsilon$ for background turbulence should have amounted to $\sim 2 - 3 \cdot 10^{-6} m^2 / s^3$ (Soloviev and Lukas, 2006). Hence, turbulence in the wake was stronger than beyond it.

Properties of flows possessing helicity can be explained by the fact that the latter effectively reduces the action of nonlinear processes responsible for energy transfer and redistribution between various scales. Indeed, writing Navier-Stokes equation for a vortex $\boldsymbol{\omega} = curl\mathbf{v}$ (in incompressible case)

$$\frac{\partial \vec{\omega}}{\partial t} = \nabla \times (\vec{v} \times \vec{\omega}) + \nu \Delta \vec{\omega} \tag{1}$$

and a relation for Lamb's vector $[\vec{v} \times \vec{\omega}]$

$$[\vec{v} \times \vec{\omega}]^2 = \vec{v}^2 \vec{\omega}^2 - (\vec{v} \cdot \vec{\omega})^2, \tag{2}$$

we can see that with growing helicity, the nonlinear term in Eq. (1) decreases and becomes zero for Beltrami flow, where $\boldsymbol{\omega} = \gamma \mathbf{v}$ ($\gamma$ being a certain constant). The same reasoning can be applied to the turbulent part of the velocity field, substituting mean velocities in Eq. (2) with fluctuational ones and averaging over the ensemble. However, as known (taking turbulence into account), the nonlinear part of Eq. (1) is connected with turbulent viscosity, so that the helical term leads to its decrease. Various studies of the properties of helical turbulence (see, for example, Belian et al., 1994, 1998; Branover et al., 1999 and references therein) demonstrate that non-zero helicity leads to a decrease in the energy flux from large to small scales.

Turbulent energy redistribution between small and large scales at nonzero helicity depends on the ratio of helicity modulus to turbulence energy (Moiseev et al., 1983a, b) and, as shown in this work, helical turbulence can lead to the generation (or stabilization) of large-scale vortical structures.

Thus, the main properties of helical turbulence are non-Kolmogorov's spectral dependence of energy density $E(k)$ (where $k$ is wave vector); turbulent viscosity decrease; energy transfer from small- to large-scale motion – inverse energy transfer, and a corresponding decrease in the turbulent energy transfer from large to small scales.

Regarding the case of helical turbulence representing a basically 3D-phenomenon (although its anisotropy can be high), we emphasize that the inverse energy transfer is its important property. The inverse energy transfer often acquires the character of an exponentially fast generation of a large-scale structure due to the motion instability. One of the most important results of the development of the helical turbulence model is the understanding of the fact that the 3D-character of motion determines the topology and, finally, intrinsic properties of turbulence. Thus, we proceed to discussing turbulent wake properties connected with the presence of nonzero mean helicity.

## Wake expansion in field experiments

To evaluate properties of turbulent viscosity in a turbulent wake, we examine the dependence of its width on the distance to the ship. An asymptotic relation for the expansion of the width of the turbulent wake aft a self-propelled ship with zero axial net momentum was predicted in 1957 by Birkhoff and Zarantonello, but only recently it has been verified for full-scale vessels in field experiments (Peltzer et al., 1992, Milgram et al., 1993a). Note two sets of performed experiments:

(1) an initial set of measurements of waves and a few surface tension measurements in and near the wakes of commercial vessels, and

(2) more comprehensive measurements in and near the wakes of several U.S. Navy vessels.

In addition, radar and background meteorological measurements were also carried out. Waves were measured with specially designed resistance transducers having a high accuracy and high signal-to-noise ratio for the range of frequencies that included waves associated with the Bragg backscattering of radars (Milgram et al, 1993a). Surface tension distributions calibrated in a wide range from 44.5 to 73 mN/m were measured by spreading oils (Peltzer et al., 1992). Extensive data on surfactants distributions across ship wakes were obtained in 1989 Field Experiment (Peltzer et al., 1992). Wake widths $\tilde{w}$ were obtained at 3 or 4 distances $x$ aft of two Navy ships, a destroyer and a frigate, at the velocities from 6.2 m/s (12 knots) to 12.9 m/s (25 knots) corresponding to the range of the Froude numbers Fr = $V_s/(gL)^{1/2}$=0.156 – 0.366, where $V_s$ and L are ship velocity and length, g – gravitational acceleration. The surface tension distribution and SAR aircraft image intensity across a wake of 5 min age were obtained approximately at the same distance 3577 m aft the ship (a destroyer at 12.9 m/s on 28/01/89) (see Milgram et al., 1993a, Fig. 11).

The widths of the wake detected in these experiments using surfactants distribution and SAR images (especially in C-band) averaged over 100 m along the wake and 12 m across the wake are evidently rather close. Milgram et al. (1993a, b) have concluded the following:

1. Reduced radar return leading to a dark centerline wake image is undoubtedly associated with reduced short-wave energy in the centerline ship wake.

2. Both turbulence and concentration of surfactants in longitudinal bands caused by the passage of a ship play a certain role in attenuating short waves. The dependence $\tilde{W}/B$ vs. x/L obtained by Milgram et al. (1993a) reads

$$\frac{\tilde{W}}{B} = \frac{\tilde{W}_0}{B}\left(\frac{x}{L}\right)^n \qquad (3)$$

where $\tilde{W}_0$ is a parameter characterizing the width of a self-similar far wake at a distance reduced to x = L aft a ship, B and L are the width and length of a ship. The dependence was approximated in log-log coordinates by a slope n = 1/5 for eight wakes (Milgram et al., 1993a).

The slopes $n$ decrease for both the destroyer and the frigate, and characteristic widths of the wake $\tilde{W}_0$/B increase with growing Fr. The widths of the wakes differ significantly (by 50%) in the experiments where Fr numbers varied two-fold for both the destroyer and the frigate. Furthermore, the frigate wakes $\tilde{W}_0$ normalized to B are approximately 30% larger than the destroyer wakes for every velocity used in the experiments. This also points to a dependence on Fr numbers, since Fr values for the frigate are less than those for the destroyer at the same velocity.

We have processed the same data independently for each wake. The slopes for seven wakes of eight are evidently less than 1/5, which is obtained for only one wake. The mean value of $n$ for the destroyer reads 0.153 and 0.133 for the frigate, that is, the slope is close to 1/7. We include the Fr number into the equation in order to reveal its effect on the wake width

$$\frac{\tilde{W}}{B} = \frac{\tilde{W}_0}{B}\left(Fr^2\,\frac{x}{L}\right)^n \tag{4}$$

That results in evident grouping of experimental data shown in Fig. 4. Note that the definition of $\tilde{W}_0$ is changed in this case, and it becomes a parameter characterizing the width of a self-similar far wake if $Fr^2\,\frac{x}{L} = 1$ aft a ship.

The values of n and $\tilde{W}_0$/B in Equation (4) for eight different wakes are presented in Table 1, where $\tilde{W}_0$/B values for each ship are close and the values for the frigate (FF) are larger than for the destroyer (DDG) by about 5-10%. Possible reasons of this slight scattering are different parameters of the hulls or/and wind conditions.

**Table 1.** Characteristics of destroyer and frigate wakes

| Run | Velocity m/s | Fr DDG | n DDG | $\tilde{W}_0$/B DDG | Fr FF | n FF | $\tilde{W}_0$/B FF | $\dfrac{\tilde{W}_0/B_{DDG}}{\tilde{W}_0/B_{FF}}$ |
|-----|-----|------|------|------|------|------|------|------|
| 1 | 6.2 | 0.18 | 0.21 | 0.83 | 0.16 | 0.16 | 0.92 | 0.90 |
| 2 | 12.9 | 0.37 | 0.17 | 0.86 | 0.32 | 0.1 | 0.92 | 0.93 |
| 3 | 9.3 | 0.27 | 0.12 | 0.85 | 0.23 | 0.13 | 0.94 | 0.90 |
| 4 | 12.9 | 0.37 | 0.15 | 0.82 | 0.32 | 0.15 | 0.86 | 0.95 |

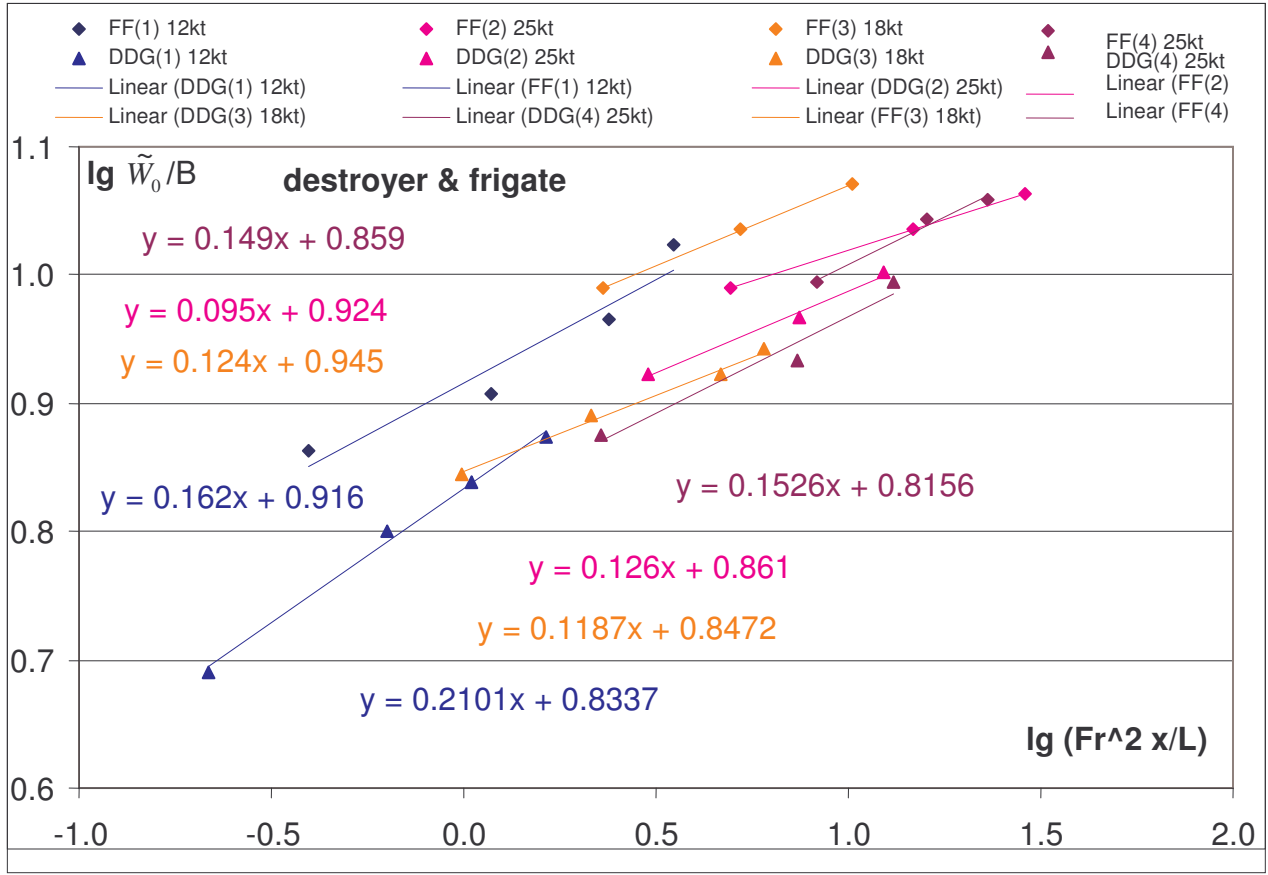

Fig.4. Dependence of the wake width $\tilde{W}/B$ on the distance aft a ship x/L and the Froude number.

**Discussion of effective turbulent viscosity in the wake**

The value of n = 1/5 in Eq. (1) suggested for experimental data by Milgram et al. (1993a) corresponds to the theoretical result obtained for the model of a wake of a self-propelled axisymmetric submerged body (Tennekes, Lumley, 1990; Birkhoff, Zarantello, 1957). According to Tennekes and Lumley (1990), the value n = 1/5 should be obtained for wakes of axisymmetric submerged bodies in the approximation of a constant turbulent viscosity over the wake cross-section.

In order to improve the description of a swirled wake, where the transverse component of the velocity field should be taken into account besides the longitudinal and radial ones, we can use a universal transverse profile for the entire wake. Because of this, equation $U\dfrac{\partial U}{\partial x}+\left\langle v\dfrac{\partial u}{\partial r}\right\rangle=0$ (where $U$ and $u$ are mean and turbulent longitudinal velocity components, v - turbulent component of the radial velocity; see also (4.1.19) in (Tennekes & Lumley, 1990)) is not valid any more. A longitudinal derivative of pressure appears in the right-hand part of this equation (Loytsansky, 1970) because of the radial component of the pressure gradient. This derivative balances centrifugal forces arising due to the wake swirl in the stationary case.

We assume that the longitudinal pressure gradient component weakly affects the universal character of the profile in certain limited sections of the wake, i.e., we introduce a piecewise universality of the wake. Here the total momentum flux through the wake cross-section is conserved (Loytsansky, 1970):

$$\int_{-\infty}^{\infty} r(p+\rho U^2)dr = const = J_0$$

and comprises the local pressure. In this section, the dependences of the characteristic width $l$ and velocity defect $(U-U_0)$ on the distance $x$, $x^n$ and $x^{n-1}$, respectively, obtained by Tennekes & Lumley (1990) also hold true. On the other hand, the conservation of the principal moment of momentum transfer through the jet cross-section in case of a swirling flow (Loytsansky, 1970) arises side by side with the conservation of the total momentum flux:

$$\rho\int_{-\infty}^{\infty} r^2 UW dr = const = L_0 \qquad\qquad (5)$$

where $W$ is the transverse velocity component. The moment $L_0$ is constant along the wake and serves as a measure of the wake swirling.

In the case of solid rotation with $W \propto r$, $L_0 \cong const\int_{-\infty}^{\infty} r^2(U-U_0)dr$, which leads to the value $n=\dfrac{1}{5}$ after de-dimensionalization. However, a wake of a ship is far from solid rotation.

We assume that, to the first approximation, $W \approx \beta(r)curl_\varphi(\mathbf{U}) = -\beta(r)\dfrac{\partial U}{\partial r}$ (where $\mathbf{U}=\{U,V,W\}$ is the mean velocity field in the wake; the pseudo-scalar $\beta \cong \dfrac{W^2}{W\cdot curl_\varphi\bar{\mathbf{U}}}$

and in stationary piecewise universality case weakly depends on $x$ ). Then it follows from (5) that

$$L_0 \cong -\rho \int_{-\infty}^{\infty} \beta(r) r^2 U \frac{\partial U}{\partial r} dr, \qquad (6)$$

and for the dependence $l \propto x^n$ to hold true and not to diverge from $\beta(r)$ at $r=0$, the condition $\dfrac{d(\beta(r) r^2)}{dr} = r^{\gamma}$ should be valid in the piecewise universality approximation. This leads to the dependence $\beta(r) \propto r^{\gamma-1}$, and here we can write after integrating by parts

$$L_0 \approx const \int_{-\infty}^{\infty} r^{\gamma} (U - U_0) dr, \qquad (7)$$

where $U_0$ is the longitudinal velocity value beyond the wake.

Therefore, to obtain the experimental value of the parameter $n$, we arrive at the equation

$$n = \frac{1}{2 + \gamma} \qquad (8)$$

by de - dimensionalizing equation (7).

The magnitude $\beta(r)$ is a pseudo-scalar and should change its sign at $-r$ substitution for $r$. Hence, the parameter $\gamma$ should assume only the values for which $(-r)^{\gamma-1} = -r^{\gamma-1}$. Experimental $n \approx \frac{1}{6} \div \frac{1}{8}$ values (see Table 1) correspond, according to (8), to $\gamma$ values within the interval from 4 to 6.  Thus, $\gamma = 5$ corresponds to average value of experimental $n \approx \frac{1}{7}$.

Effective turbulent diffusivity $v_t^{\exp} = v_t^{eff} = \dfrac{\langle u_x u_r \rangle}{\dfrac{\partial U_x}{\partial r}}$ determined from the experimental data on the ratio of the measured component $\{xr\}$ of Reynolds stresses tensor to the radial derivative of the longitudinal velocity is not constant over the wake cross-section and decreases from the centre to the vortex periphery (Patel, Sadra, 1990). Such behavior of the effective turbulent viscosity can be due to the fact that in the general case of local homogeneous isotropic turbulence with violated mirror symmetry, the correlation velocity tensor includes, besides the symmetrical part, an asymmetric one produced by mean helicity. The symmetric part is responsible for the dissipation, i.e. can be identified with turbulent viscosity (see, e.g., Krause, Redler, 1980; Monin, Yaglom,1996). The asymmetric part can compensate, under certain conditions, the symmetric part decreasing the effective turbulent viscosity (Belian et al., 1994, 1998).

To analyze turbulent viscosity properties in a wake cross-section, we make use of the equation derived by Moiseev et al. (1983) for mean velocity (vorticity) in the presence of nonzero mean turbulent helicity:

$$\frac{\partial \mathbf{\Omega}}{\partial t} - rot(\mathbf{U} \times \mathbf{\Omega}) - curl(\alpha\mathbf{\Omega}) = curl \nabla \nu_t \nabla \mathbf{U} \tag{9}$$

where $\mathbf{\Omega} = curl \mathbf{U}$, $\alpha \approx \tau He$, $\tau$ - effective relaxation time, $\nu_t$ - turbulent viscosity defined by the symmetrical part of the turbulent velocity field correlator, i.e. $\nu_t \approx \tau E$, where $E$ is the mean energy of turbulence (see also Branover et al., 1999, and references therein). If $\alpha$ and $\nu_t$ are constants, we obtain the well-known equation

$$\frac{\partial \mathbf{\Omega}}{\partial t} - rot \ (\mathbf{U} \times \mathbf{\Omega}) - \alpha \, curl \ \mathbf{\Omega} \ = \nu_t \Delta \mathbf{\Omega} \tag{10}$$

describing vortex generation. This equation in a somewhat expanded form was used for the study of large-scale vortical structures generation in various conditions including planetary atmospheres (see Branover et al., 1999; Ivanov et al., 1996 and references therein). Actually, a positive parameter of instability $\delta = i\omega = |\alpha|k - \nu_t k^2$ (where $k$ is wave vector) for modes with $k < \frac{|\alpha|}{\nu_t}$ can be obtained with Fourier transformation of (10). At the same time, modes with $k > \frac{|\alpha|}{\nu_t}$ attenuate, but the rate of this attenuation decreases with the growth of $|\alpha|$. On the other hand, $\nu_t k$ corresponds to the operator $\nu_t \nabla$ in Eq. (9), which determines turbulent viscosity. Thus, the presence of nonzero mean helicity leads to the reduction of the effective turbulence viscosity (Belyan et al., 1994, 1998; Golbraikh et al., 1998).

To demonstrate this assertion, we rewrite the expression under *curl* operator in Eq. (9) as

$$\frac{\partial U}{\partial t} - (\mathbf{U} \times \mathbf{\Omega})_x = \alpha \Omega_x + \nabla \nu_t \nabla U \tag{11}$$

where in the above approximations in the cylindrical coordinates

$\nabla \nu_t \nabla U = \dfrac{1}{r}\dfrac{\partial}{\partial r}\left(r\nu_t \dfrac{\partial}{\partial r}U\right)$ and $\Omega_x = \dfrac{1}{r}\dfrac{\partial}{\partial r}rW = -\dfrac{1}{r}\dfrac{\partial}{\partial r}\left(r\beta(r)\dfrac{\partial U}{\partial r}\right)$. Then we can rewrite the

right-hand side of (11) as

$$(\alpha\beta(r) - \nu_t)\frac{\partial U_x}{\partial r} = \nu_t(\frac{\alpha}{\nu_t}\beta(r) - 1)\frac{\partial U_x}{\partial r} \tag{12}$$

If the signs of the mean turbulent helicity and large-scale helicity are the same, the effective turbulent viscosity $\nu_t^{eff} = \nu_t\left(1 - \dfrac{\alpha}{\nu_t}\beta\right) = \nu_t\left(1 - \dfrac{\beta}{\beta_t}\right)$ (where $\beta_t = \dfrac{\nu_t}{\alpha}$) decreases. At the same time, since $\beta$ grows from the center to the periphery, $\nu_t^{eff}$ should decrease from the center to the periphery, which is observed experimentally (Patel, Sadra, 1990) and, in principle, agrees with the results of Hoekstra and Ligtelijn (1991).

Now we estimate the orders of magnitude of the parameters $\nu_t$, $\alpha$ and $\beta$ for experimental data (Patel, Sadra, 1990). We introduce a radius-averaged value $\tilde{\beta} = \dfrac{1}{R}\int_0^R \beta(r)dr$, where $R$ is the effective vortex size. The model dimensions were as follows: L = 3.048 m, B = 0.305 m and D = 0.19 m.

At a moderate distance from the ship, we can estimate $\tilde{\beta}$ as $\tilde{\beta} \approx \dfrac{W}{\dfrac{\partial U}{\partial r}}$ and $\beta_t = \dfrac{\nu_t}{\alpha}$ from the difference between parameters at the levels $Z/D = 0$ and $Z/D = 0.5$ (size of the vortex is on the order of half-depth of the wake). Since $\tilde{\beta} \approx 0.1D$ and $\beta_t \approx 0.1B$ and $\nu_t \approx \dfrac{\nu_t^{eff}}{(1 - \dfrac{D}{B})} \approx 1.6\nu_t^{eff}$, which points to about 1.5-fold viscosity decrease from the vortex center to its periphery.

Thus, the generated helical turbulence leads to a decrease in the effective turbulent viscosity, which is reflected in a lower increase in the wake width with the distance than the previous model predicts.

## Conclusions

Helical turbulence has not been practically applied to oceanic phenomena. It is partially a result of limited experimental data (as compared, e.g., to the atmosphere) on oceanic turbulent fields. We make an attempt, based on available experimental data, to show that helical turbulence generated in turbulent ship wakes can essentially affect dynamic characteristics of the wake. In particular, such effect should be observed in spectra of turbulent velocity fluctuations in the wake, which become steeper ($E(k) \propto k^{-7/3}$) than Kolmogorov's spectra. At the same time, effective turbulent viscosity in the wake should decrease, which results in slowing-down of the wake widening with the distance from the ship. Besides, it should lead to a far wake stabilization, which is possible in the presence of helical turbulence due to the permanent inverse energy cascade from small to large scales, side by side with the regular energy flux from large to small scales. In the present study, we do not discuss energy sources maintaining turbulence in the wake, but it is noteworthy that since helicity mixes up velocity components, any large-scale energy input into turbulence in the far wake maintains turbulence on the whole. However, it is noteworthy that they can include processes arising against the background of wind waves generation, changes in the surface tension coefficient in the wake as compared with the surroundings, interaction of wake with Kelvin's waves, as well as pressure fluctuations connected with the ship and transmitted over large distances, and some others to be discussed in conclusion.

Generative properties of helical turbulence have been examined in numerous studies both in conductive and non-conductive media. Various external impacts connected with rotation, shear flows, thermal convection, etc. have been taken into account in these studies. For example, Moiseev et al. (1983) have shown using equations (9) or (10) that in $L > \dfrac{\nu_t}{|\alpha|}$ scales instability arises leading to the generation (enhancement) of coherent vortical structures of typhoon type. We believe that similar processes of generation (stabilization) also occur in the case of longitudinal vortices enhancement in a ship wake.

This follows from the work of Novikov (1996) who has shown that the change in the velocity circulation even in the presence of a free surface and surfactant is determined only by the circulation of viscous force $\frac{\partial \Gamma}{\partial t} = \oint \nabla \nu_t^{eff} \nabla U_i \delta r_i$ ($i$ varies from 1 to 3). In our work, we have introduced an effective viscosity instead of kinematic one into Novikov's expression for velocity circulation. Therefore, if turbulence possesses helical properties, and the effective turbulent viscosity can be reduced, the vorticity will be maintained long enough.

We have shown that the widening of a turbulent ship wake can be related to the helicity magnitude while the effective viscosity $\nu_t^{eff}$ decreases from the center of the wake to its periphery. Accordingly, the rate of its widening decreases while the life-time of the ship wake increases.